\documentclass[twocolumn,trackchanges]{aastex701}

\usepackage{orcidlink}
\usepackage[version=4]{mhchem}

\begin{document}

\title{Sub-Neptunes Are Drier Than They Seem: Rethinking the Origins of Water-Rich Worlds}

\shortauthors{Werlen et al. }
\correspondingauthor{Aaron Werlen}
\shorttitle{Rethinking the Origins of Water-Rich Worlds}

\author[orcid=0009-0005-1133-7586, sname='Werlen']{Aaron Werlen}
\affiliation{Institute for Particle Physics and Astrophysics, ETH Zurich, CH-8093 Zurich, Switzerland}
\email[show]{awerlen@ethz.ch}  

\author[orcid=0000-0001-6110-4610, sname='Dorn']{Caroline Dorn}
\affiliation{Institute for Particle Physics and Astrophysics, ETH Zurich, CH-8093 Zurich, Switzerland}
\email{dornc@ethz.ch}

\author[orcid=0000-0002-9020-7309, sname="Burn"]{Remo Burn}
\affiliation{Max Planck Institute for Astronomy, 69117 Heidelberg, Germany}
\email{burn@mpia.de}

\author[orcid=0000-0002-0298-8089, sname='Schlichting']{Hilke E. Schlichting}
\affiliation{Department of Earth, Planetary, and Space Sciences, University of California, Los Angeles, CA 90095, USA}
\email{hilke@ucla.edu}

\author[orcid=0000-0002-0632-4407, sname='Grimm']{Simon L. Grimm}
\affiliation{Institute for Particle Physics and Astrophysics, ETH Zurich, CH-8093 Zurich, Switzerland}
\affiliation{Department of Astrophysics, University of Zurich, CH-8057 Zurich, Switzerland}
\email{sigrimm@ethz.ch}

\author[orcid=0000-0002-1299-0801, sname='Young']{Edward D. Young}
\affiliation{Department of Earth, Planetary, and Space Sciences, University of California, Los Angeles, CA 90095, USA}
\email{eyoung@epss.ucla.edu}

\begin{abstract}
Recent claims of biosignature gases in sub-Neptune atmospheres have renewed interest in water-rich sub-Neptunes with surface oceans, often referred to as Hycean planets. These planets are hypothesized to form beyond the snow line, accreting large amounts of \ce{H2O} ($>$10~wt\%) before migrating inward. However, current interior models often neglect chemical equilibration between primordial atmospheres and molten interiors. Here, we compute global chemical equilibrium states for a synthetic population of sub-Neptunes with magma oceans. Although many initially accrete 5–30~wt\% water, interior–atmosphere interactions destroy most of it, reducing final \ce{H2O} mass fractions to below 1.5~wt\%. As a result, none meet the threshold for Hycean planets. Despite that, we find \ce{H2O}-dominated atmospheres exclusively on planets that accreted the least ice. These planets form inside the snow line, are depleted in carbon and hydrogen, and develop small envelopes with envelope mass fractions below 1\%, dominated by endogenic water. In contrast, planets formed beyond the snow line accrete more volatiles, but their water is largely converted to \ce{H2} gas or sequestered into the interior, resulting in low atmospheric \ce{H2O} mass fractions. Most \ce{H2O}-rich envelopes are also fully miscible with \ce{H2}, making a separate water layer unlikely. Our results challenge the conventional link between ice accretion and water-rich atmospheres, showing instead that \ce{H2O}-dominated envelopes emerge through chemical equilibration in hydrogen-poor planets formed inside the snow line.
\end{abstract}



\section{Introduction}\label{sec:intro}

The idea of water-rich ‘Hycean’ exoplanets has gained significant attention following JWST observations of the sub-Neptune K2-18b, which suggests potential biosignatures \citep{madhusudhan_habitability_2021, madhusudhan_carbon-bearing_2023,madhusudhan_new_2025}. However, both the interpretation of these spectral features and the classification of K2-18b as a Hycean planet remain highly debated, with alternative scenarios suggesting that it may instead be a gas-rich sub-Neptune with a magma ocean, which would preclude the presence of a habitable water ocean \citep[e.g.][]{shorttle_distinguishing_2024, wogan_jwst_2024, schmidt_comprehensive_2025, jordan_planetary_2025, taylor_are_2025, welbanks_challenges_2025}.

The idea that some sub-Neptunes might host extensive water-rich volatile layers builds on earlier theoretical work by \citet{kuchner_volatile-rich_2003} and \citet{leger_new_2004}, who proposed that planets that formed beyond the snow line and migrated inwards could retain significant amounts of volatile ices such as \ce{H2O}. Combined formation and evolution models suggest that some sub-Neptunes can be water worlds (up to 50\% water by mass) that migrated inwards from outside the water snowline \citep{mousis_irradiated_2020, venturini_nature_2020, chakrabarty_where_2024, burn_radius_2024} where silicate-to-ice ratios of 1:1 are reached \citep{lodders_solar_2003}. The presence of close-in water-rich sub-Neptunes has been predicted by many studies \citep[e.g.,][]{mordasini_extrasolar_2009, alibert_formation_2017, burn_new_2021, bitsch_dry_2021, venturini_fading_2024}. In particular, \citet{bitsch_dry_2021} predict that, in the absence of giant planets, sub-Neptunes can migrate inward while retaining several tens of weight percents of water, forming water-rich envelopes even in the inner disk. 

However, one fundamental limitation of these formation studies is that they neglect any chemical coupling between volatiles and deep interiors. Studies like \citet{kite_water_2021}, \citet{schlichting_chemical_2022}, \citet{misener_atmospheres_2023}, \citet{luo_interior_2024}, \citet{seo_role_2024}, \citet{tian_atmospheric_2024}, \citet{rogers_most_2024} and \citet{werlen_atmospheric_2025} showed that chemical interactions between magma ocean and atmosphere can significantly modify the bulk and atmospheric composition of sub-Neptunes.

While many observed evolved sub-Neptunes may currently harbor deep magma oceans \citep{kite_atmosphere_2020}, all super-Earths and sub-Neptunes are thought to have commonly experienced this state early in their evolutionary histories. During formation, super-Earths and sub-Neptunes can accrete a few weight percent of hydrogen–helium gas \citep[e.g.][]{lee_cool_2015, ginzburg_super-earth_2016}, which when present as an outer envelope is enough to trap heat acquired during the accretion process. The resulting temperatures at the atmosphere–magma ocean interface are several thousands of kelvin or more \citep{ginzburg_super-earth_2016,misener_importance_2022,young_phase_2024} and the optically thick hydrogen envelopes can delay magma ocean solidification for up to gigayear timescales \citep[e.g.][]{lopez_understanding_2014, ginzburg_super-earth_2016}. During these prolonged magma ocean phases, atmosphere–interior interactions can significantly alter the composition and evolution of sub-Neptunes \citep[e.g.,][]{ginzburg_super-earth_2016, chachan_role_2018, kite_superabundance_2019, kite_water_2021, schlichting_chemical_2022,misener_importance_2022,tian_atmospheric_2024, seo_role_2024, burn_water-rich_2024, werlen_atmospheric_2025}. 

In this study, we analyze the influence of atmosphere–magma ocean equilibration on the water content of young sub-Neptune exoplanets. We compute global chemical equilibrium states for a large sample of synthetic planets generated by a planetary population synthesis model. These planets span a range of bulk compositions, envelope mass fractions, and orbital distances. The population is extracted from the simulations just after disk dispersal and hence represents a young and hot population for which magma oceans are a plausible scenario. We then compare the equilibrated water mass fractions to the accreted values predicted by the population synthesis model, enabling an assessment of how the bulk water inventory is modified by interior–atmosphere interactions. In addition to bulk properties, we extract the envelope \ce{H2O} mass fraction for each planet from the equilibrium composition and analyze how it varies with orbital distance and bulk properties. Finally, we analyze the miscibility of \ce{H2O }and \ce{H2} in envelopes for our population.

This paper is structured as follows. In Section \ref{sec:Method}, we describe the global chemical equilibrium model and outline the planetary population synthesis framework that provides the initial conditions for our analysis. We also detail the simulation setup and population selection. In Section \ref{sec:Results}, we present the distribution of water mass fractions across the planetary population, identify differences between initial and equilibrated states, and analyze trends in envelope \ce{H2O} mass fractions with orbital distance and bulk properties. In Section \ref{sec:Discussion}, we summarize our results, explore the implications for planet formation, and compare our findings with previous studies. Finally, in Section \ref{sec:Conclusions}, we summarize the main conclusions of this work.

\section{Methods}\label{sec:Method}

\subsection{Chemical Thermodynamics}\label{sec:chemical_network}

We use the global chemical equilibrium network described in \citet{werlen_atmospheric_2025}, which is based on the framework developed by \citet{schlichting_chemical_2022}, with the key addition of carbon partitioning in the metal phase. The network includes 19 chemical reactions involving 26 phase components distributed across three coexisting phases: metal, silicate, and gas (the envelope).

We determine the abundances of the 26 phase components, along with the total number of moles in each phase, by simultaneously enforcing chemical equilibrium, mass balance, and elemental conservation in each phase. Our numerical scheme is adapted from the approach developed by \citet{schlichting_chemical_2022} and extended by Grimm et al. (in preparation) with substantial improvements in computational efficiency. For a full description of the equilibrium formulation, including the reaction network and governing equations, we refer the reader to Appendix~\ref{sec:appendix_equilibrium}, as well as \citet{schlichting_chemical_2022} and \citet{werlen_atmospheric_2025}.

As in \citet{schlichting_chemical_2022}, \citet{young_phase_2024} and \citet{werlen_atmospheric_2025}, we define the astrophysical core as comprising both metal and silicate phases. This definition reflects the possibility that the two phases are not separated by a distinct core–mantle boundary, due to considerable sequestration of light elements into the metal. Experimental studies support the solubility of light elements in metallic phases under high-pressure conditions \citep{hirao_compression_2004, terasaki_hydrogen_2009, tagawa_experimental_2021}, a result that is further supported by ab initio calculations \citep{li_earths_2020, luo_interior_2024}. The accumulation of light elements in the metal can lower its density to values comparable to those of silicate, potentially inhibiting gravitational segregation \citep{young_phase_2024}.

\subsection{New Generation Planetary Population Synthesis (NGPPS) model}\label{ch:NGPPS}

We use the Generation III Bern model, also known as the New Generation Planetary Population Synthesis (NGPPS) model, which combines planet formation and evolution in a self-consistent framework. For a detailed description, we refer the reader to \citet{emsenhuber_new_2021} for the model description, \citet{emsenhuber_planetary_2021} for the simulation set-up and \citet{burn_water-rich_2024, burn_radius_2024} for the recent improvements on evolutionary calculations. 

The formation module simulates the evolution of an $\alpha$-viscous, accreting gas disk, the dynamics of solids within the disk, and the concurrent accretion of solids and gas by growing protoplanets. Our simulations assume one Solar mass for the central star. The models include gas-driven migration and gravitational interactions between multiple embryos, both of which are important for the emergence of water-rich sub-Neptunes. The gas disk is modeled as a one-dimensional, axisymmetric structure whose time evolution is governed by the advection–diffusion equation, with additional sink terms for planetary accretion and photoevaporation.

Planet formation begins with 50 embryos of initial mass $10^{-2}$~M$_\oplus$, distributed with a logarithmic probability density up to 40~AU. Each embryo accretes small (600\,m) planetesimals using a statistical treatment for the evolution of the planetesimals' dynamical state. The thermal structure and gas accretion of each planet’s envelope is determined by solving the one-dimensional, hydrostatic, internal structure equations, allowing the model to track both mass and radius throughout all phases of evolution. The solution of the envelope structure also determines the atmosphere-magma ocean interface (AMOI) temperature under the important approximations that the full gravitational potential energy of the accreted solids is released at this location, that the upward energy flux (i.e. the luminosity) remains radially constant, and that the dust opacity in the upper, radiative part of the envelope is reduced due to grain growth and settling \citep{mordasini_grain_2014}. This choice has the important effect that gas accretion proceeds in a regime intermediate to the dust-free and dusty cases explored in \citet{lee_cool_2015}.

The model uses prescriptions for Type I and Type II migration, eccentricity and inclination damping of non-circular orbits from non-isothermal simulations to calculate additional force terms. Those are added to gravitational forces in the \texttt{mercury} N-body code \citep{chambers_hybrid_1999} used to evolve orbits for 20~Myr.

The disk’s initial chemical composition is calculated using a chemical equilibrium model \citep{thiabaud_stellar_2014} for refractory species, assuming solar elemental abundances \citep{lodders_solar_2003}, and ratios of volatile species \citep[i.e. \ce{H2O}, \ce{CO}, \ce{CO2}, \ce{CH4}, \ce{CH3OH}, \ce{NH3}, \ce{N2}, \ce{H2S},][]{marboeuf_planetesimals_2014} informed from observations of the interstellar medium and cometary compositions. At each radial location, the disk's initial pressure and temperature is used to determine which species condense \citep{marboeuf_planetesimals_2014} and become available for accretion. These solids are incorporated into planetary embryos via planetesimal accretion, defining the initial bulk composition of forming planets. We note that in the set-up used here, refractory minerals include significant amounts of oxygen but we excluded refractory carbon carriers (e.g. graphite, refractory organics), which implies that bulk C/O ratios can reach zero in interiors within the \ce{CH3OH} snowline, which is the first snowline of carbon-bearing molecules. We exclude refractory carbon from our model based on a scenario in which interstellar grains, responsible for carrying roughly 50\% of the cosmic carbon budget in refractory form \citep[e.g.][]{whittet_dust_2022}, undergo significant heating during the earliest stages of star formation. This thermal processing is thought to destroy much of the refractory carbon in the earliest stages of star formation \citep{bhandare_mixing_2024}. Such a scenario is consistent with the low carbon abundances observed in meteorites and on Earth \citep[e.g.][]{alexander_measuring_2017}, as well as with evidence from white dwarf pollution studies \citep[e.g.][]{jura_extrasolar_2014}.

\subsection{Coupling the NGPPS and the chemical network}\label{coupling}

The NGPPS and the global thermodynamics code are coupled through the atomic bulk composition, total planetary mass, and AMOI temperature. The population is extracted from the simulations at the time of disk dispersal, when the planets are still very young and hot.

The atomic molar abundances are derived directly from the NGPPS output. We include seven key elements in the thermodynamic calculations: H, O, C, Mg, Si, Fe, and Na. Species containing elements not included in these calculations such as Al, N, S, Ni, and He are omitted. Notably, helium always accounts for 24\% of the envelope mass, meaning a substantial fraction of mass may be excluded depending on the atmospheric mass fraction. For the pressure calculation, the full planetary mass is used, as all species, reactive or not, contribute to the total pressure.

Since Na is not included in the NGPPS elemental inventory, we set its molar abundance to a fixed value at least four orders of magnitude lower than the least abundant element to avoid numerical issues without influencing the solution. These small amounts of Na are still included in the equilibrium calculation, but the resulting Na-bearing phase species remain extremely minor due to the very low initial abundance.

\subsection{Simulation Setup \& Population} \label{population}

From the birth NGPPS population, we select planets with total masses between 2 and 15 Earth masses and AMOI temperatures between 1900 and 4000~K, where a global magma ocean is expected to exist. The upper temperature limit of 4000~K is chosen to avoid excessive extrapolation of the thermodynamic data, which are calibrated based on experimental measurements at lower temperatures. The temperature at the Silicate–Metal Equilibrium (SME) boundary is assumed to be 500~K above the AMOI temperature to reflect internal thermal gradients. The SME boundary is not treated as a core–mantle boundary, but rather as a mean depth of silicate–metal equilibration.

The dataset contains 4267 planets with masses between 2 and 15 Earth masses. We first exclude 386 planets with AMOI temperatures below 1700~K and 3456 planets above 4000~K, leaving 436 planets within the relevant thermal range. Of these, 330 have envelope masses greater than $10^{-3}$ Earth masses, which we assume to be a minimal requirement for sustaining a magma ocean over prolonged timescales. Planets with lower envelope masses are likely unable to maintain a magma ocean over geologic periods. Among the 330 remaining planets, we exclude 68 planets with bulk C/O ratios below $10^{-3}$. These low C/O values result from the absence of refractory carbon in the NGPPS population. Moreover, such values are inconsistent with white dwarf pollution studies, which suggest that planetary building blocks generally contain C/O ratio above $10^{-3}$ \citep{wilson_carbon_2016}. Finally, three additional planets are ejected during their dynamical evolution, resulting in a final sample of 248 planets.

Although our sample is smaller than the full population, it remains representative, as it spans a wide range of bulk elemental properties. For example, within our sample, the C/O ratio ranges from $10^{-3}$ to 0.13, the Mg/Si ratio remains close to unity, and the O/H ratio spans from 0.18 to 15.65. These distributions are consistent with those of the full NGPPS population, indicating that the selected sample captures the key compositional diversity of the broader population.

While we exclude planets with AMOI temperatures above 4000~K to avoid extrapolating beyond the calibration range of our thermodynamic model, these hotter planets would also host magma oceans and undergo chemical equilibration. Our preliminary tests suggest that increasing the AMOI temperature has little effect on net \ce{H2O} production. We therefore expect our main conclusions regarding \ce{H2O} retention to extend to the hotter regime. Nonetheless, additional high-temperature thermodynamic measurements are needed to fully verify this expectation. Recent ab initio studies also suggest that 4000~K likely represents a physical upper bound for the AMOI temperature, as non-ideal mixing between hydrogen and silicates at higher temperatures leads to miscibility between \ce{MgSiO3} and \ce{H2} \citep{young_phase_2024,stixrude_core-envelope_2025}.

\section{Results} \label{sec:Results}

We compute the chemical equilibrium state for the 248 planets described in Section~\ref{population}. Convergence is assessed by evaluating the squared sum of the residuals from the equilibrium, mass conservation, and elemental sum constraints as described in Appendix~\ref{sec:appendix_equilibrium}. A model is considered converged if this sum falls below $10^{-7}$, in order to avoid including potentially non-converged solutions that can arise due to the presence of numerous local minima, as reported by \citet{schlichting_chemical_2022}. Out of the 248 models, 233 converged, corresponding to a convergence rate of 94\%. While including the remaining solutions does not break any of the trends observed in our results, we retain the conservative threshold to minimize the risk of reporting solutions corresponding to local rather than global minima.

We define the equilibrated \ce{H2O} as the sum of \ce{H2O} in gas and in silicates. In reality, at low \ce{H2O} fugacities, water dissolves in silicate melts predominantly as hydroxyl (\ce{OH}-) groups. However, at high \ce{H2O} fugacities ($\sim$ 2000 bars), molecular \ce{H2O} becomes the dominant species in the melt \citep{stolper_speciation_1982,berndt_combined_2002}. As dissolved water can in principle outgas during the evolution and replenish the envelope, all water present in silicates is accounted for in the total equilibrated \ce{H2O}.

\subsection{\ce{H2O} Accreted vs. H$_\textit{2}$O Equilibrated}

\begin{figure}
    \centering
    \includegraphics{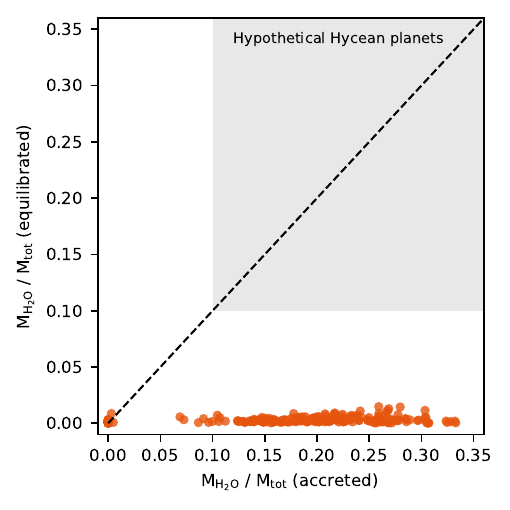}
    \caption{Comparison between the accreted and equilibrated water mass fractions of sub-Neptunes. The black dashed line indicates the 1:1 correlation; in the absence of chemistry, all planets would lie along this line. The grey shaded region denotes the 10–90 wt\% water mass fraction range proposed for Hycean planets by \citet{madhusudhan_habitability_2021}. All planets in our sample exhibit significant water depletion following equilibration and fall well below the Hycean threshold.} \label{fig:H2O_comparison}
\end{figure}

Figure~\ref{fig:H2O_comparison} compares the accreted and equilibrated \ce{H2O} mass fractions for all planets in our sample. After equilibration, all planets exhibit low \ce{H2O} total mass fractions, typically below 1.5 wt\%, regardless of how much water they initially accreted. Even planets that began with up to 30~wt\% \ce{H2O} lose the vast majority through interior–atmosphere chemical redistribution. The gray box indicates the \ce{H2O} mass fraction range proposed for Hycean planets (10–90~wt\%) by \citet{madhusudhan_habitability_2021}; all of our planets fall well below this threshold, including those that accreted large amounts of water.

\subsection{Envelope H$_\textit{2}$O Content and Formation Location}

\begin{figure}
    \centering
    \includegraphics{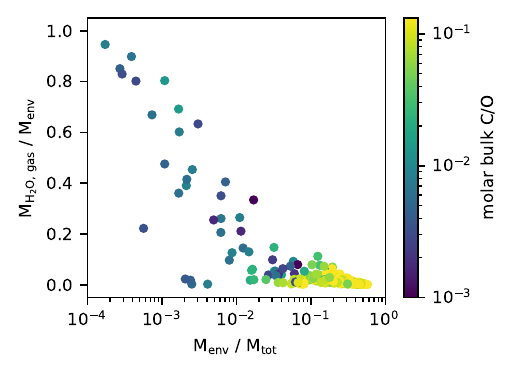}
    \caption{Envelope \ce{H2O} mass fraction as a function of envelope mass fraction, using the same dataset as in Figure~\ref{fig:H2O_comparison}. The colorbar indicates the molar bulk C/O ratio. Planets with low envelope mass fractions tend to retain a higher proportion of \ce{H2O} in the gas phase. Moreover, planets with low C/O ratios consistently show higher \ce{H2O} content compared to their carbon-rich counterparts.}
    \label{fig:H2O_comparison_env}
\end{figure}

Figure~\ref{fig:H2O_comparison_env} shows the envelope \ce{H2O} mass fraction as a function of atmosphere mass fraction, with the colorbar denoting the molar bulk C/O ratio. A clear trend emerges: planets with atmosphere mass fractions below 10\% exhibit significantly higher envelope \ce{H2O} mass fractions, often reaching values between 10\% and 90\%. In contrast, planets with more massive envelopes typically have \ce{H2O} mass fractions below 10\%. This inverse correlation is an effect of dilution: in hydrogen-poor envelopes, even small absolute amounts of \ce{H2O} can dominate the mass budget of the envelope. As the envelope mass—and thus the hydrogen content—increases, the \ce{H2O} mass fraction decreases accordingly. This holds true, because the total \ce{H2O} mass fractions for gas-poor and gas-rich planets are very similar. This underlines that the water mass fraction of 1.5~wt\% (Figure \ref{fig:H2O_comparison}) is a universal limit with little dependency on planet mass or gas mass fraction. 

\begin{figure*}
    \centering
    \includegraphics[width=1\textwidth]{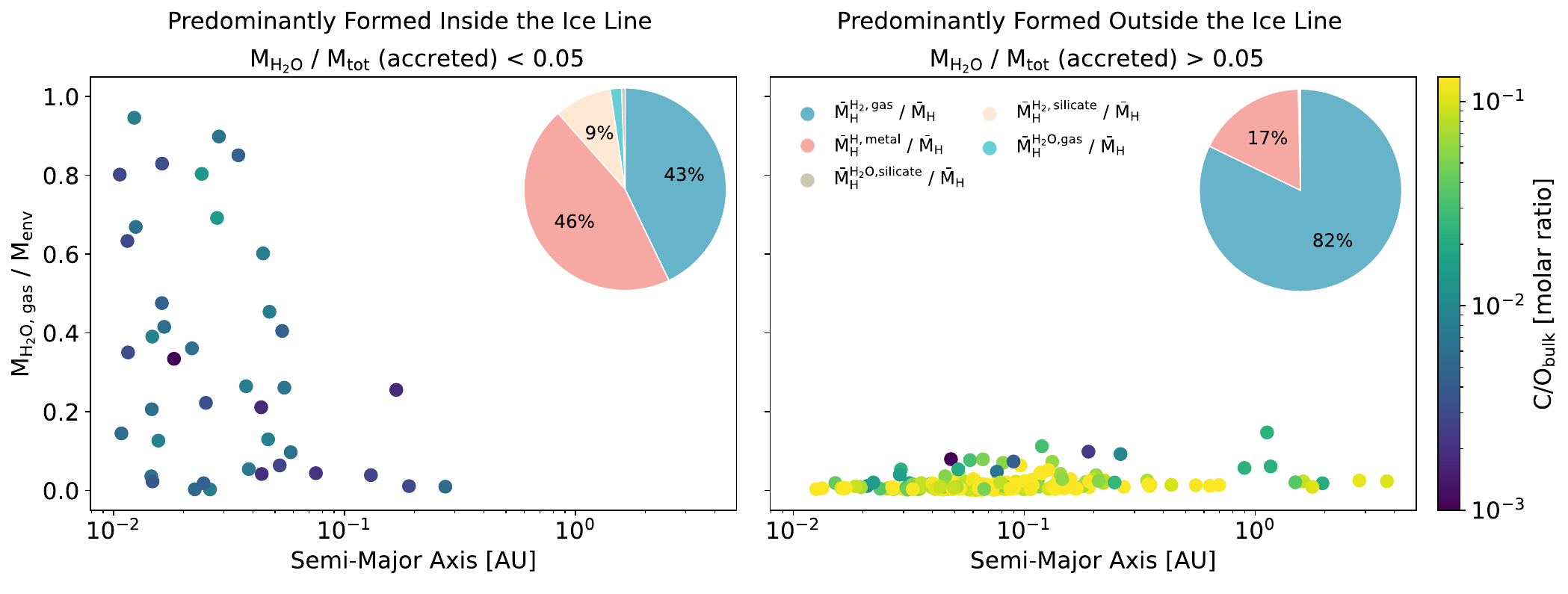}
    \caption{Envelope \ce{H2O} mass fraction as a function of semi-major axis. The same dataset as in Figure~\ref{fig:H2O_comparison} is shown. The left panel shows planets that predominantly formed inside the water ice line; the right panel shows those that formed outside. Classification is based on the accreted \ce{H2O} mass fraction, with a threshold set at 5\% of the total planetary mass. The colorbar indicates the molar bulk C/O ratio. Planets formed inside the ice line are systematically depleted in carbon due to the lack of volatile ice accretion and exhibit higher envelope \ce{H2O} mass fractions. In contrast, planets formed beyond the ice line retain lower \ce{H2O} content despite higher bulk volatile abundances. Each pie chart shows the mean mass fraction of hydrogen in \ce{H2} (gas), H (metal), \ce{H2} (silicate), \ce{H2O} (gas), and \ce{H2O} (silicate), normalized to the total mean hydrogen inventory for each population. Only components contributing more than 5\% are labeled. Planets that formed beyond the ice line store most hydrogen as \ce{H2} (gas) and H (metal), while those that formed inside the ice line retain a larger share of hydrogen in H (metal) \ce{H2} (silicate) and \ce{H2O} (gas + silicate).}
    \label{fig:orbital_distr.}
\end{figure*}

\begin{figure}
    \centering
    \includegraphics{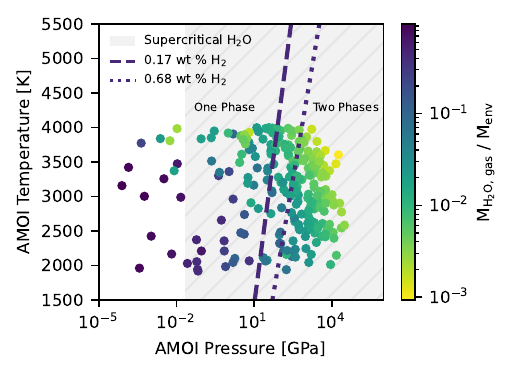}
    \caption{Atmosphere–magma ocean interface (AMOI) temperature versus AMOI pressure. The same dataset as in Figure~\ref{fig:H2O_comparison} is shown. The colorbar indicates the mass fraction of \ce{H2O} in the envelope. The grey dashed region marks the domain where pure \ce{H2O} exists in the supercritical state \citep{yang_subcritical_2007}. The dashed and dotted lines indicate the \ce{H2}–\ce{H2O} solvus at $0.68$~wt\% \ce{H2} and $0.17$~wt\% \ce{H2}, respectively. The latter corresponds to the critical composition where the solvus in pressure–composition space reaches its minimum \citep{gupta_miscibility_2025}. Planets to the left of the solvus fall within the one-phase regime, where \ce{H2} and \ce{H2O} are fully miscible. Planets to the right lie in the two-phase regime, where the two species coexist as separate phases. Most planets with high \ce{H2O} gas mass fractions lie within the one-phase regime.}
    \label{fig:solvus}
\end{figure}

Outliers to this trend (Figure~\ref{fig:H2O_comparison_env} bottom left) can be explained by planets with high AMOI temperatures. In these cases, elevated temperatures enhance the evaporation of species such as Mg, Si, and SiO, thereby reducing the relative \ce{H2O} mass fraction. More generally, planets with low atmosphere mass fractions also tend to exhibit lower bulk C/O ratios, suggesting they accreted less carbon-bearing material. These compositional differences reflect variations in formation location.

To explore this connection, Figure~\ref{fig:orbital_distr.} shows the envelope \ce{H2O} mass fraction as a function of orbital distance. Because planets undergo Type I and Type II migration during their formation, the orbital distances shown correspond to their final locations after disk dispersal. To infer formation regions, we divide the sample based on the accreted \ce{H2O} mass fraction: planets with less than 5~wt\% are assumed to have formed inside the ice line, while those with more likely formed beyond it, where volatile ices were more abundant. Planets that predominantly formed beyond the ice line also accreted substantial amounts of CO$_2$ and CH$_3$OH, leading to higher bulk C/O ratios. 

On average, planets that formed beyond the ice line initially accreted 23~wt\% of hydrogen-rich gas, compared to only 2~wt\% for those that formed inside it. 
The propensity of planets in the outer disk to accrete more hydrogen can be traced to the opacities in our model planets. Cooling leads to accretion. Temperature dependent dust opacities influence the rate of cooling above the radiative-convective boundary and therefore the masses of the accreted envelopes. In our models, relatively lower opacities in the outer disk result from lower equilibrium temperatures and less dust (in agreement with the analytic analysis by \citet{lee_cool_2015}, see also \citet{coleman_situ_2017} for a numerical exploration). If we were to use greater opacities, cooling would become uniformly less efficient, and the dependence of envelope mass on semi-major axis would be virtually eliminated. Our calculated planets with reduced dust opacities lie intermediate to the cases explored by \citet{lee_cool_2015}. We note that the boil-off phase could alter this scaling \citep{ginzburg_super-earth_2016}, but we assume that chemical equilibrium is reached during the disk stage prior to loss processes \citep{rogers_under_2024}. Despite their enhanced hydrogen and carbon budgets, these planets exhibit low envelope \ce{H2O} mass fractions, as much of the hydrogen is redistributed during equilibration into \ce{H2} gas, \ce{CH4} gas, and hydrogen dissolved in the metallic phase.

To further examine how hydrogen is partitioned, Figure~\ref{fig:orbital_distr.} shows pie charts displaying the mean hydrogen distribution across volatile reservoirs for the two populations. Planets formed beyond the ice line store most of their hydrogen as \ce{H2} gas and H in metal, while those formed inside the ice line retain more hydrogen in the form of \ce{H2O} (gas and melt) and \ce{H2} in silicate. Because these inner planets accreted much less total hydrogen, even modest amounts of \ce{H2O} can dominate the envelope composition, leading to a broader diversity in envelope \ce{H2O} content. The apparent shift in partitioning between the two populations arises primarily from dilution: while the absolute abundances of H-bearing trace species like \ce{H2O} are comparable, they constitute a larger fraction of the hydrogen budget in inner planets because these planets accreted less hydrogen during formation.

\subsection{Miscibility of H\texorpdfstring{$_\textit{2}$}{2}–H\texorpdfstring{$_\textit{2}$}{2}O in the Envelope}

At first glance, some of the planets in our sample, particularly those with high \ce{H2O} gas mass fractions, might appear to be candidates for shallow surface oceans. However, this interpretation is misleading. In our model, the thermodynamic conditions at the AMOI rule out the existence of distinct, stable liquid water phases.

Figure~\ref{fig:solvus} shows AMOI temperature against AMOI pressure for all planets in our sample. The colorbar indicates the \ce{H2O} mass fraction in the gas phase of the envelope. The grey dashed area marks the pressure–temperature domain where pure \ce{H2O} exists in the supercritical state \citep{yang_subcritical_2007}. The dashed and dotted lines represent the \ce{H2}–\ce{H2O} solvus from \citet{gupta_miscibility_2025} for 0.68~wt\% \ce{H2} and 0.17~wt\% \ce{H2}, in a binary \ce{H2}–\ce{H2O} mixture. The 0.17~wt\% \ce{H2} curve corresponds to the critical composition at which the solvus reaches its pressure minimum. The solvus defines the boundary between one-phase and two-phase regions in composition–pressure–temperature space, marking the conditions under which hydrogen and water become immiscible and separate into distinct phases. Planets located to the left of this solvus fall within the one-phase regime, where hydrogen and water are fully miscible. Planets to the right may, in principle, exhibit phase separation at the AMOI in case their composition and temperature fall within the narrow two-phase window.

In practice, all planets in our sample with high \ce{H2O} envelope mass fractions lie well within the one-phase regime, where hydrogen and water form a single supercritical and superionic fluid. Even among those located to the right of the solvus, the AMOI temperatures exceed 1700~K and 100~GPa, well above the critical temperature and pressure of \ce{H2O}, resulting in supercritical rather than liquid water. Moreover, due to the steep positive $dT/dP$ slope of the solvus, any potential two-phase region at depth would rapidly transition to full miscibility with altitude as pressure decreases.

Note that the solvus curves shown here are derived for a binary \ce{H2}-\ce{H2O} system and are therefore only valid under the assumption that these two species dominate the envelope composition. In planets with very low \ce{H2O} mass fractions, other species such as \ce{CH4}, \ce{CO2}, \ce{CO}, and \ce{SiO} can become increasingly abundant. In such cases, the assumption of a binary mixture breaks down, and the solvus curves presented here no longer provide reliable predictions of phase behaviors.

\section{Discussion} \label{sec:Discussion}

Our results, which focus on the initial (birth) population of sub-Neptunes with magma oceans, suggest that their water mass fractions are not primarily set by the accretion of icy pebbles during formation but by chemical equilibration between the primordial atmosphere and the molten interior. None of the planets in our model, regardless of their initial \ce{H2O} content, retain more than 1.5~wt\% water after chemical equilibration. This excludes the high water mass fractions (10–90~wt\%) invoked by Hycean-world scenarios \citep{madhusudhan_habitability_2021}, even for planets that initially accreted up to 30\% \ce{H2O} by mass. These findings are consistent with recent studies suggesting that only a small amount of water can be produced or retained endogenously in sub-Neptunes and super-Earths \citep[e.g.,][]{rogers_most_2024}.

Despite these low bulk water inventories, a subset of planets develop small \ce{H2O}-rich envelopes. These are H- and C-poor planets that formed inside the ice line, where they accreted no ices and less hydrogen compared to planets that formed outside the ice line. Their small, hydrogen-depleted envelopes are more readily dominated by minor volatile species, resulting in high \ce{H2O} mass fractions despite a low absolute water content. This outcome is broadly consistent with the predictions by \citet{kite_water_2021}, who propose \ce{H2O}-dominated atmospheres with surface pressures between $10^{-5}$ and $0.2$ GPa for planets with orbital periods shorter than 100 days. In \cite{kite_water_2021} these atmospheres form through efficient oxidation of a hydrogen-rich primordial atmosphere by a magma ocean, followed by atmospheric escape. We see the same trend in our results, although we do not account for atmospheric escape. 

In addition to neglecting atmospheric escape, we only analyze the birth population and do not account for fractionated mass loss, which can alter the composition of the envelope over long timescales by a preferential loss of hydrogen (e.g. \citet{cherubim_strong_2024}; Valatsou et al. (in preparation)).

Our model also naturally reproduces key compositional trends reported in previous studies, such as the strong dependence of atmospheric C/O and O/H ratios on envelope mass fraction, as demonstrated by \citet{seo_role_2024} and \citet{werlen_atmospheric_2025}.

\begin{figure}
    \centering
    \includegraphics{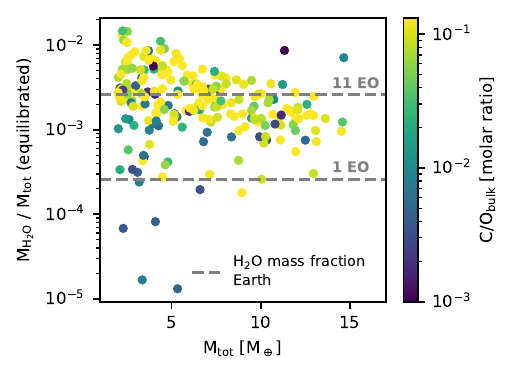}
    \caption{Equilibrated \ce{H2O} mass fraction as a function of total planet mass, using the same dataset as in Figure~\ref{fig:H2O_comparison}. The colorbar indicates the molar bulk C/O ratio. The grey dashed lines mark the lower and upper estimate of Earth's water mass fraction, corresponding to 1 and 11 Earth oceans (EO) respectively. This estimate agrees well with our results, suggesting that the \ce{H2O} content of sub-Neptunes is broadly consistent with Earth's upper bound. A negative correlation is visible between planet mass and \ce{H2O} content for planets with high C/O ratios.}
    \label{fig:Mp}
\end{figure}

The low \ce{H2O} mass fractions in our study are broadly consistent with upper estimates for Earth's interior water content. Geochemical and geophysical models suggest that up to $\sim$10 Earth oceans (EO) of water, where 1~EO corresponds to $1.4 \times 10^{21}$kg, may be stored in the mantle \citep[e.g.,][]{hirschmann_water_2006,albarede_volatile_2009,genda_origin_2008,elkins-tanton_ranges_2008,hamano_emergence_2013,nakagawa_global-scale_2017}. As shown in Figure~\ref{fig:Mp}, our post-equilibration water mass fractions span $10^{-3}$ to 1.5~wt\%, with a mean of 0.27~wt\%, overlapping Earth's estimated range (0.025–0.25~wt\%). The negative correlation between planet mass and upper \ce{H2O} mass fraction in high C/O planets arises from how oxygen is partitioned during interior–atmosphere equilibration. Although more oxygen is available in higher-mass planets, a larger fraction is sequestered into the metal phase, leaving a nearly constant oxygen abundance with planet mass in the silicate and gas phases. As a result, the absolute \ce{H2O} mass remains roughly unchanged across the population, while the relative \ce{H2O} mass fraction decreases with increasing planet mass.

Interior water storage further complicates the relationship between accreted and observable water content. \citet{luo_interior_2024} show that substantial amounts of water can be sequestered into a planet’s metallic core based on pressure-dependent metal–silicate partitioning at pressures up to 1000~GPa. Their model includes pressure effects, which we omit for reasons discussed in \citet{schlichting_chemical_2022} and \citet{werlen_atmospheric_2025}. Including such pressure dependence would likely refine our estimates but not change our main findings. If H and O in the metal phase were interpreted as bound \ce{H2O}, the total water inventory would increase significantly. However, this water would remain locked deep in the interior, inaccessible to the surface or observable atmosphere.

Crucially, even planets with \ce{H2O}-dominated envelopes in our model cannot sustain surface oceans. Most are firmly within the one-phase \ce{H2}–\ce{H2O} regime, where hydrogen and water are fully miscible. Even those near the critical composition remain above 1700~K at the AMOI, preventing the presence of liquid water. 

Although our results reflect birth conditions, the long-term evolution of sub-Neptunes may further limit the potential for surface ocean formation. Recent work by \citet{rogers_road_2025} suggests that small, \ce{H2O}-dominated envelopes cool inefficiently over gigayear timescales due to the high opacity of water. Moreover, planets closer to 0.1~AU with Bond albedos of $>$0.3 reach equilibrium temperatures above 650~K, which is well above the \ce{H2}–\ce{H2O} solvus at pressures below 0.5~GPa \citep{gupta_miscibility_2025}. Together, these findings imply that even the small water-rich envelopes in our sample are unlikely to cool sufficiently to cross into the two-phase regime.

While our model captures a broad range of redox and partitioning processes, it includes only ferrous iron (\ce{Fe^{2+}}) and does not account for reactions between ferric iron (\ce{Fe^{3+}}) species and \ce{H2}. \citet{horn_reaction_2023} have shown that additional \ce{H2O} can be produced via the reaction \ce{Fe2O3 + 4H2 = 2FeH + 3H2O} under high-pressure conditions. This yields three moles of \ce{H2O} per mole of \ce{Fe2O3}, compared to just one mole of \ce{H2O} per mole of \ce{FeO} in our model that occurs by the reaction \ce{FeO + H2 = H2O + Fe}, which is a linear combination of the basis vector reactions R16 – R17 – R3 – \ce{1/2}R7 (see Appendix~\ref{sec:appendix_equilibrium}). However, the overall contribution to the \ce{H2O} budget is likely limited, as the abundance of ferric iron relative to total iron oxides is typically low at relevant oxygen fugacities ($<$ 0.1; \citet{hirschmann_magma_2022}). Based on this ratio, the maximum increase in \ce{H2O} from ferric iron reactions would remain below 20\%. Nonetheless, explicitly incorporating \ce{Fe^{3+}} into future models could improve the accuracy of \ce{H2O} budget estimates.

Beyond the equilibrium modeling presented here, several broader caveats and future directions warrant consideration. From the population synthesis side, our analysis is based on a subset of synthetic planets and may not capture the full diversity of formation histories or disk conditions. In particular, variations in pebble dynamics can influence the radial distribution of volatiles and refractory elements, potentially altering the bulk compositions of forming planets \citep[e.g.][]{morbidelli_dynamics_2012,johansen_pebble_2021,castrejon_effect_2024}. Similarly, the presence of refractory carbon phases, which is absent from our current models, could significantly impact the C/O ratio, affecting water production and atmospheric chemistry \citep[e.g.][]{werlen_atmospheric_2025}. Addressing these complexities will require future studies that more tightly couple dynamical formation pathways with detailed compositional modeling. 

Furthermore, an important next step will be to link interior–atmosphere equilibrium models of the planet birth population with evolution models in order to compare to observational data. While such an effort lies beyond the scope of this study, our results provide a necessary foundation for interpreting sub-Neptunes in the context of interior processes and formation environments.

\section{Conclusion}\label{sec:Conclusions}

Using a representative population of super-Earths and sub-Neptunes at birth, we find that planets with magma oceans retain less than 1.5~wt\% water by mass after undergoing interior–atmosphere chemical equilibration—even when initially accreting up to 30~wt\% water. This rules out the formation of Hycean-like planets with deep surface oceans and suggests that Earth’s water content may be a typical outcome of early planetary evolution.

Counterintuitively, the planets with the most water-rich atmospheres are not those that accreted the most ice, but those that are depleted in hydrogen and carbon. These planets typically form inside the ice line and accrete less volatile-rich material. While some retain significant atmospheric \ce{H2O}, the high-temperature miscibility of water and hydrogen likely prevents the presence of surface liquid water—even on these comparatively water-rich worlds.

These findings challenge the classical link between ice-rich formation and water-rich atmospheres, and instead highlight the dominant role of magma ocean–atmosphere equilibration in shaping planetary compositions. This has broad implications for theories of planet formation and volatile evolution, as well as for interpreting exoplanet atmospheres in the era of JWST, ELT, ARIEL, HWO, and LIFE. It also informs atmospheric composition priors in interior characterization of transiting planets observed by Kepler, TESS, CHEOPS, and PLATO with RV or TTV constraints.


\section*{Acknowledgements}

A.W. gratefully acknowledges travel support from the European Research Council (ERC) Synergy Geoastronomy grant under grant number 101166936. C.D acknowledges support from the Swiss National Science Foundation under grant TMSGI2\_211313. R.B. acknowledges the financial support from DFG under Germany’s Excellence Strategy EXC 2181/1-390900948, Exploratory project EP 8.4 (the Heidelberg STRUCTURES Excellence Cluster). H.E.S gratefully acknowledges support from NASA under grant number 80NSSC18K0828. E.D.Y. acknowledges support from NASA grant number 80NSSC21K0477 issued through the Emerging Worlds program. This work has been carried out within the framework of the NCCR PlanetS supported by the Swiss National Science Foundation under grant 51NF40\_205606. We thank the anonymous reviewer for their insightful comments, which greatly helped to improve this study. We acknowledge the use of large language models (LLMs), including ChatGPT, to improve the grammar, clarity, and readability of the manuscript.

\section*{ORCID iDs}

\noindent 
Aaron Werlen \orcidlink{0009-0005-1133-7586} \href{https://orcid.org/0009-0005-1133-7586}{0009-0005-1133-7586} \\
Caroline Dorn \orcidlink{0000-0001-6110-4610} \href{https://orcid.org/0000-0001-6110-4610}{0000-0001-6110-4610} \\
Remo Burn \orcidlink{0000-0002-9020-7309} \href{https://orcid.org/0000-0002-9020-7309}{0000-0002-9020-7309} \\
Hilke E. Schlichting \orcidlink{0000-0002-0298-8089} \href{https://orcid.org/0000-0002-0298-8089}{0000-0002-0298-8089} \\
Simon L. Grimm \orcidlink{0000-0002-0632-4407} \href{https://orcid.org/0000-0002-0632-4407}{0000-0002-0632-4407} \\
Edward D. Young \orcidlink{0000-0002-1299-0801} \href{https://orcid.org/0000-0002-1299-0801}{0000-0002-1299-0801}

\appendix
\twocolumngrid

\section{Chemical Network}\label{sec:appendix_equilibrium}

The chemical network is constructed following the approach in \cite{schlichting_chemical_2022}. Like in \cite{werlen_atmospheric_2025}, we add carbon as a new species to the metal phase. In total, the system has 19 independent reactions involving 26 phase components. Chemical exchange is allowed both within and between the core, who is composed of a silicate and metal phase as well as the gas.

Reactions within the silicate phase are:

\begin{equation}
    \ce{Na2SiO3_{,silicate} \rightleftharpoons Na2O_{silicate} + SiO2_{,silicate}} \tag{R1}
\end{equation}

\begin{equation}
    \ce{MgSiO3_{,silicate} \rightleftharpoons MgO_{silicate} + SiO2_{,silicate}} \tag{R2}
\end{equation}

\begin{equation}
    \ce{FeO_{silicate} + 1/2Si_{metal} \rightleftharpoons Fe_{metal} + 1/2SiO2_{,silicate}} \tag{R3}
\end{equation}

\begin{equation}
    \ce{FeSiO3_{,silicate} \rightleftharpoons FeO_{silicate} + SiO2_{,silicate}} \tag{R4}
\end{equation}

Reactions between the metal and the silicate phase include:

\begin{equation}
    \ce{O_{metal} + 1/2Si_{metal}} \rightleftharpoons \ce{1/2SiO2_{,silicate}} \tag{R5}
\end{equation}

\begin{equation}
    \ce{2H_{metal} \rightleftharpoons H2_{,silicate}} \tag{R6}
\end{equation}

\begin{equation}
    \ce{Si_{metal} + 2H2O_{silicate} \rightleftharpoons SiO2_{,silicate} + 2H2_{,silicate}} \tag{R7}
\end{equation}

\begin{equation}\label{Carbonreaction}
    \ce{C_{metal} + O_{metal} \rightleftharpoons CO_{silicate}} \tag{R8}
\end{equation}

Reactions among gas-phase species are:

\begin{equation}
    \ce{CO_{gas} + 1/2O2_{,gas} \rightleftharpoons CO2_{,gas}} \tag{R9}
\end{equation}

\begin{equation}
    \ce{CH4_{,gas} + 1/2O2_{,gas}} \rightleftharpoons \ce{2H2_{,gas} + CO_{gas}} \tag{R10}
\end{equation}

\begin{equation}
    \ce{H2_{gas} + 1/2O2_{,gas} \rightleftharpoons H2O_{gas}} \tag{R11}
\end{equation}

Reactions representing magma ocean–atmosphere exchange are:

\begin{equation}
    \ce{FeO_{silicate} \rightleftharpoons Fe_{gas} + 1/2O2_{,gas}} \tag{R12}
\end{equation}

\begin{equation}
    \ce{MgO_{silicate} \rightleftharpoons Mg_{gas} + 1/2O2_{,gas}} \tag{R13}
\end{equation}

\begin{equation}
    \ce{SiO2_{,silicate} \rightleftharpoons SiO_{gas} + 1/2O2_{,gas}} \tag{R14}
\end{equation}

\begin{equation}
    \ce{Na2O_{silicate}} \rightleftharpoons \ce{2Na_{gas} + 1/2O2_{,gas}} \tag{R15}
\end{equation}

\begin{equation}
    \ce{H2_{,silicate} \rightleftharpoons H2_{,gas}} \tag{R16}
\end{equation}

\begin{equation}
    \ce{H2O_{silicate} \rightleftharpoons H2O_{gas}} \tag{R17}
\end{equation}

\begin{equation}
    \ce{CO_{silicate} \rightleftharpoons CO_{gas}} \tag{R18}
\end{equation}

\begin{equation}
    \ce{CO2_{,silicate} \rightleftharpoons CO2_{,gas}} \tag{R19}
\end{equation}

Note that any additional reaction that can be formed as a linear combination of these basis reactions is also permitted. The reaction space is therefore not limited to the expressions explicitly listed above.

Chemical equilibrium among the 19 reactions is determined by solving the following condition for the molar fractions $x_i$ of all participating species:

\begin{equation}\label{eq:chemical_equilibrium}
    \sum_i \nu_i \ln x_i + \left[\frac{\Delta \hat{G}^\circ_{\text{rxn}}}{RT} + \sum_g \nu_g \ln(P/P^\circ)\right] = 0,
\end{equation}

\noindent where $x_i$ denotes the mole fraction of species $i$ within its respective phase, and $\nu_i$ are the associated stoichiometric coefficients. The term $\Delta \hat{G}^\circ_\text{rxn}$ represents the standard Gibbs free energy change of the reaction, $R$ is the ideal gas constant, $T$ is the temperature, $P$ is the pressure at the atmosphere–magma ocean interface (AMOI), and $P^\circ$ is the reference pressure, set here to 1 bar. The summation over $g$ accounts for gas-phase species only, which introduces the explicit pressure dependence.

As previously described in \citet{young_earth_2023}, we account for the non-ideal mixing between Si and O in the metal phase by replacing the mole fractions $x_i$ of Si and O in Equation~\ref{eq:chemical_equilibrium} with their respective activities $a_i$. Following \citet{badro_core_2015}, the activity is defined as $a_i = \gamma_i x_i$, where $\gamma_i$ is the activity coefficient.

The activity coefficient of Si is given by:

\begin{align}
&\ln \gamma_{\mathrm{Si}} = -6.65 \frac{1873 \,\mathrm{K}}{T} - 12.41 \frac{1873 \,\mathrm{K}}{T} \ln(1 - x_{\mathrm{Si}}) \\
&\quad + 5 \frac{1873 \,\mathrm{K}}{T} x_{\mathrm{O}} \left( 1 + \frac{\ln(1 - x_{\mathrm{O}})}{x_{\mathrm{O}}} - \frac{1}{1 - x_{\mathrm{Si}}} \right) \nonumber \\
&\quad - 5 \frac{1873 \,\mathrm{K}}{T} x_{\mathrm{O}}^2 x_{\mathrm{Si}} \left( \frac{1}{1 - x_{\mathrm{Si}}} + \frac{1}{1 - x_{\mathrm{O}}} \right. \nonumber \\
& \quad \left. + \frac{x_{\mathrm{Si}}}{2(1 - x_{\mathrm{Si}})^2} - 1 \right) \nonumber.
\end{align}

\noindent The activity coefficient of O is given by:

\begin{align}
&\ln \gamma_{\mathrm{O}} = 4.29 - \frac{16500 \,\mathrm{K}}{T} + \frac{16500 \,\mathrm{K}}{T} \ln(1 - x_{\mathrm{O}}) \\
&\quad + 5 \frac{1873 \,\mathrm{K}}{T} x_{\mathrm{Si}} \left( 1 + \frac{\ln(1 - x_{\mathrm{Si}})}{x_{\mathrm{Si}}} - \frac{1}{1 - x_{\mathrm{O}}} \right) \nonumber \\
&\quad - 5 \frac{1873 \,\mathrm{K}}{T} x_{\mathrm{Si}}^2 x_{\mathrm{O}} \left( \frac{1}{1 - x_{\mathrm{O}}} + \frac{1}{1 - x_{\mathrm{Si}}} \right. \nonumber \\
&\quad \left. + \frac{x_{\mathrm{O}}}{2(1 - x_{\mathrm{O}})^2} - 1 \right) \nonumber.
\end{align}

To account for non-ideal mixing between O and C in the metal phase, we use the activity coefficient expression from \citet{fischer_carbon_2020}:

\begin{equation}
    \ln \gamma_{\mathrm{C}} = -2.303 \cdot 19.5 \ln(1- x_{\mathrm{O}})
\end{equation}

In addition to the equilibrium conditions, we applied three normalization constraints to ensure that the mole fractions in each phase sum to unity:

\begin{equation}
    1 - \sum_i x_{i,k} = 0,
\end{equation}

\noindent where $x_{i,k}$ is the mole fraction of species $i$ in phase $k$.

Elemental conservation is enforced through seven additional constraints—one for each conserved element—requiring that:

\begin{equation}
    n_s - \sum_i \sum_k n_{s,i,k} x_{i,k} N_k = 0,
\end{equation}

\noindent where $n_s$ is the total abundance of element $s$, $n_{s,i,k}$ is the number of atoms of element $s$ in species $i$ of phase $k$, $x_{i,k}$ is again the mole fraction of species $i$ in that phase, and $N_k$ is the total number of moles in phase $k$. We treat both $x_{i,k}$ and $N_k$ as free variables across the three considered phases: gas, silicate, and metal. The AMOI pressure is internally determined via the calculated mean molecular weight of the atmosphere.

To solve for global equilibrium, we adopt the numerical method developed by \citet{schlichting_chemical_2022}, modified to significantly reduce computational cost from roughly 30 minutes to a few seconds on a single CPU core. Specific algorithmic improvements are detailed in Grimm et al. (in preparation).

We use the same thermodynamic data as described in the appendix of \cite{schlichting_chemical_2022}. For the carbon partitioning reaction (\ref{Carbonreaction}), which was first introduced in \cite{werlen_atmospheric_2025}, we adopt the carbon partitioning coefficients from \cite{blanchard_metalsilicate_2022} for the 2+ valence state of carbon. Pressure corrections are omitted, as they were found to be negligible. The Gibbs free energy of reaction is calculated as:

\begin{equation}
    \Delta \hat{G}^\circ_{\text{rxn,R8}} = 2.303 \left(0.3 + \frac{3882 \,\mathrm{K}}{T} \right) + \Delta \hat{G}^{\circ,\text{metal}}_{\text{O}},
\end{equation}

\noindent where $T$ is the temperature and $\Delta \hat{G}^{\circ,\text{metal}}_{\text{O}}$ is the standard Gibbs free energy of formation of oxygen in the metal phase.

\newpage

\bibliography{references}
\bibliographystyle{aasjournal}



\end{document}